\documentstyle[amsmath,aps,epsfig]{revtex}
\renewcommand{\Im}{{\mathcal I}m}
\renewcommand{\Re}{{\mathcal R}e}
\newcommand{\er}{{\mathbf{r}}}
\newcommand{\pp}{{\mathbf{p}}}
\newcommand{\vu}{{\mathbf{v}}}
\newcommand{\Vu}{{\mathbf{V}}}
\newcommand{\brak}[1] {\langle #1 \rangle}
\renewcommand{\theta}{\vartheta}
\renewcommand{\phi}{\varphi}
\renewcommand{\epsilon}{\varepsilon}
\begin{document}
\title{Collisional damping of the collective oscillations of a trapped
  Fermi gas} \author{Lorenzo Vichi} \address{Dipartimento di Fisica,
  Universit\`a degli studi di Trento and Istituto Nazionale per la
  Fisica della Materia, I-38050 Povo, Italy} 
\date{June 20, 2000} 
\maketitle
\begin{abstract}
  We consider a Fermi gas confined by a harmonic trapping potential
  and we highlight the role of the Fermi--Dirac statistics by studying
  frequency and damping of collective oscillations of quadrupole type
  in the framework of the quantum Boltzmann equation, in which
  statistical corrections are taken into account in the collisional
  integral.  We are able to describe the crossover from the
  collisionless regime to the hydrodynamic one by introducing a
  temperature-dependent relaxation time $\tau_Q$. We show that, in the
  degenerate regime, the relaxation rate $\tau_Q^{-1}$ exhibits a
  temperature dependence different from the collision rate
  $\gamma$. We finally compare the collisional properties of the Fermi
  gas with the ones of the Bose gas for temperatures above the
  Bose--Einstein condensation.
\end{abstract}
\pacs{}
\section{Introduction}
Since the experimental realization of Bose--Einstein condensation in
trapped metastable atomic gases \cite{BEC} the investigation of
collective oscillations has proven to be an important tool to
characterize the behaviour of these many body
systems. Indeed, the high accuracy of frequency measurements
\cite{dip-temperatura} and the
sensitivity of collective phenomena to interaction effects
(see \cite{Giorgini} and references therein for a recent discussion of
collective excitations in condensed Bose gases at finite temperature)
makes them good candidates to check the predictions of theory
\cite{rmp}.
The quantum degenerate regime has also been recently reached in a
magnetically trapped Fermi gas \cite{Jin} and measurements of
collective oscillations might be available soon also for fermionic
species.

The aim of this paper is to discuss the temperature dependence of the
frequency and damping of the low energy collective oscillations of
a trapped Fermi gas in its normal, non superfluid phase. The
transition temperature to the superfluid phase is predicted to occur
at very low temperature \cite{Kagan,Stoof}, while the effects of quantum
degeneracy should be visible before the transition
\cite{Silvera,Butts,Schneider,Bruun}. For a discussion of the 
oscillations in the superfluid phase see \cite{Stoof,superfluid-oscill}.
Collective
oscillations can occur in the collisionless ($\omega \tau_Q \gg 1$) or
in the hydrodynamic ($\omega \tau_Q \ll 1$) 
regime. Here $\omega$ is the frequency of the oscillation and $\tau_Q$
is a characteristic relaxation time. We calculate
the relaxation time $\tau_Q$ by linearizing the Boltzmann
equation and we show that it is important to use $\tau_Q$
instead of the inverse of the collision rate $\gamma$, which measures
the time between collisions. The two times are proportional in the
classical regime, but exhibit drastically different $T$ dependences in
conditions of quantum degeneracy.
In addition to the study of the
temperature dependence of $\tau_Q$ in the Fermi case, which depicts
the role of quantum statistics, we also compare the relaxation time
of the collective oscillations with the one exhibited by a Bose
gas above the critical 
temperature $T_c$ for Bose--Einstein condensation, to highlight the
role of the bosonic statistics.

The Pauli exclusion principle dramatically changes the
collisional properties of Fermi gases in the quantum degenerate regime
and the onset of degeneracy can be 
traced down by looking at the corresponding changes of collective
oscillations. At very low temperature Fermi--Dirac statistics quenches
the number of collisions between atoms and one expects to find the gas in
the collisionless (CL) regime. At higher temperatures there is the
possibility of finding the gas in the hydrodynamic (HD) regime,
especially if the cross section is very large, as in
the case of $^6$Li. An example of such a crossover has been given in
\cite{SD} for the spin dipole oscillation.

The Fermi gas is taken
to be a two component (``spin up'' and ``spin down'') gas, to avoid
the absence of $s$-wave collisions in the sample. The Bose gas will be
instead considered to be  trapped in a single spin state.
We use a method recently developed by Gu\'ery-Odelin \emph{et al.} in
\cite{DavidFrancesca} to evaluate in relaxation time 
approximation the collisional integral. The relaxation time $\tau_Q$
is obtained through a linearization of the Boltzmann equation evaluated
with an ansatz which describes the deformations in momentum space
driven by the external oscillation. A similar technique has also been
developed by Al Khawaja \emph{et al.} \cite{AlKawaja} for the
evaluation of the 
collective excitations of a non-condensed Bose gas. An analytical and
numerical investigation of the relaxation time 
and of the collision rate in a non-condensed uniform Bose gas was done by
Lopez-Arias and Smerzi in \cite{Smerzi} while Kavoulakis \emph{et al.}
\cite{Kavoulakis} presented a microscopic calculation of the
relaxation time of a trapped non-condensed Bose gas based on the
Boltzmann equation. Our approaches lead, in fact, to the same
expression for the relaxation time.

The paper is organized as follows: in Section \ref{sec2} we briefly
describe the method of the averages for a two component Fermi gas and we
introduce the relaxation time $\tau_Q$ for the quadrupolar collective
oscillation and compare it with the collisional rate $\gamma$. 
Section~\ref{sec3} is dedicated to the comparison with bosons. 
Section~\ref{sec4} reports the results for the collective
oscillations of the Fermi gas as a function of the
temperature. Section~\ref{sec5} draws conclusions and perspectives. 

\section{Relaxation time and collision rate}
\label{sec2}

The equations describing the dynamical evolution of the quadrupole
oscillations in a two component Fermi system can be derived starting
from the Boltzmann equation 
\begin{equation} \label{BE}
\left( \frac \partial{\partial t} + \frac {\pp_1} m  \cdot
  \boldsymbol{\nabla} + 
  {\mathbf{F}}_\sigma \cdot \boldsymbol{\nabla}_{\pp_1}
\right) f_\sigma(\er,\pp_1,t)= I_{coll}[f_\sigma]
\end{equation}
for the phase space distribution function for a given spin component
$\sigma$, $f_\sigma(\er,\pp,t)$ \cite{KadanoffBaym} where
${\mathbf{F}}_\sigma= - \boldsymbol{\nabla} V^\sigma_{ext}(\er)$ is
the force originating from the external trapping potential, that we
choose to be axially symmetric and that is, in general, different for
different spin components
\begin{equation}
V^\sigma_{ext}(\er)=\frac 1 2 m \left[ \omega_{\sigma,\perp}^2 \left
  ( x^2 +y^2 \right) +
  \omega_{\sigma,z}^2 z^2 \right].
\end{equation}
The collisional integral reads
\begin{multline} \label{CollInt}
I_{coll}[f_\sigma]= \frac {\sigma_f} {4 \pi h^3
  m} \int
d^3p_2\,d\Omega\, 
|\pp_1-\pp_2| \times \\
\left[ (1 - f_\sigma(\pp_1))(1 -
  f_{-\sigma}(\pp_2))f_\sigma(\pp'_1)f_{-\sigma}(\pp'_2)-
f_\sigma(\pp_1)f_{-\sigma}(\pp_2)(1 - f_\sigma(\pp'_1))(1-
f_{-\sigma}(\pp'_2)) \right].
\end{multline}
We have used the notation $f_\sigma(\pp_i)\equiv
f_\sigma(\er,\pp_i,t)$.  The moments $\pp_1'$ and $\pp'_2$ are
constrained by energy and momentum conservation:
$p_1^2+p_2^2=p^{\prime\, 2}_1+p^{\prime\, 2}_2$ and $\pp_1+\pp_2=
\pp_1'+\pp_2'$. As the gas is at very low temperature the collisional
cross section is taken to be $\sigma_f=4\pi a^2$ with $a$ the $s$-wave
scattering length for up-down collisions. These collisions are the
dominant ones at low temperatures, as up-up and down-down collisions
occur in the $p$ channel for antisymmetry requirements \cite{Taylor}.
By neglecting $p$-wave scattering, suppressed by the centrifugal
barrier at the low energies of interest \cite{DeMarco}, we allow
interactions only between up and down particles.

The Boltzmann equation (\ref{BE}) can be used to describe the trapped
gas if one can use the semiclassical approximation and if mean field
contributions to the energy of the gas are negligible. The former
condition reads $k_BT \gg \hbar \omega_{\perp,z}$ and can be fulfilled
over a wide range of temperatures for large number of trapped
particles as the Fermi temperature, below which quantum degeneracy
effects begin to show, is defined as $T_F=(3N)^{1/3}\hbar
\omega_{ho}/k_B$ (we have introduced the geometrical average of the
trapping frequencies $\omega_{ho}=(\omega_\perp^2 \omega_z)^{1/3}$).
The condition for the semiclassical approximation is then $T/T_F \gg
(3N)^{-1/3}$ and is ensured if $N$ is sufficiently large. Mean field
effects are small the case for fermions \cite{SD} and this is also
true for the Bose gas above the critical temperature for
Bose--Einstein condensation
$T_c=[N/\zeta(3)]^{1/3}\hbar\omega_{ho}/k_B$. The reason is that the
gas is dilute and the mean field interactions, of the order of $gn(0)$
(here $g=4\pi \hbar^2 a/(2m)$ with $n(0)$ the central density), must
be compared to the kinetic energy which, above $T_c$, is of the order
of $k_BT$. The correction is typically of the order of few percent.

When working with a two component system it is convenient to use the
total distribution function $f=f_\uparrow+f_\downarrow$ and the
difference $f_-=f_\uparrow-f_\downarrow$. By summing and subtracting
the Boltzmann equation for the up and the down component one gets
\begin{equation} \label{sum}
\left( \partial_t + \frac {\pp_1} m  \cdot
  \boldsymbol{\nabla} + 
  {\mathbf{F}}_+ \cdot \boldsymbol{\nabla}_{\pp_1}
\right) f(\er,\pp_1,t)+{\mathbf{F}}_- \cdot
  \boldsymbol{\nabla}_{\pp_1}f_-(\er,\pp_1,t)=
  I_{coll}[f_\uparrow]+I_{coll}[f_\downarrow] 
\end{equation}
and
\begin{equation} \label{diff}
\left( \partial_t + \frac {\pp_1} m  \cdot
  \boldsymbol{\nabla} + 
  {\mathbf{F}}_+ \cdot \boldsymbol{\nabla}_{\pp_1}
\right) f_-(\er,\pp_1,t)+{\mathbf{F}}_- \cdot
  \boldsymbol{\nabla}_{\pp_1}f(\er,\pp_1,t)=
  I_{coll}[f_\uparrow]-I_{coll}[f_\downarrow] 
\end{equation}
where ${\mathbf{F}}_\pm=({\mathbf{F}}_\uparrow \pm
{\mathbf{F}}_\downarrow)/2$. In the right hand side of
equations~(\ref{sum})-(\ref{diff}) the collisional contributions
should be expressed in terms of $f$ and $f_-$. In the simplified case
of equal trapping potentials
$V_{ext}^\uparrow=V_{ext}^\downarrow=V_{ext}$ for the two spin
components one has ${\mathbf{F}}_-=0$. If also
$N_\uparrow=N_\downarrow$ then the difference $f_-$ is zero, at
equilibrium, for symmetry. In this work we will restrict our attention
only to in-phase
oscillations. One is then left with one equation, in which the
only unknown is the total phase space distribution function
$f=2f_\uparrow=2f_\downarrow$.

A systematic way to solve the Boltzmann equation is provided by the
Chapman--Enskog expansion \cite{CE} which allows to introduce the
viscosity and the thermal conductivity in the hydrodynamic
equations that will then exhibit damping. To actually calculate the
viscosity and the thermal conductivity it is necessary to provide some
ansatz for their functional dependence.

Starting from equation~(\ref{sum}) with $f_-=0$ we instead
derive equations for the average of generic dynamical quantities
$\chi(\er,\pp)$: 
\begin{equation}
 \partial_t \brak{\chi} - m^{-1} \brak{\pp \cdot
\boldsymbol{\nabla} \chi} - \brak{{\mathbf{F}} \cdot
\boldsymbol{\nabla} _\pp \chi}
= \sum_\sigma\brak{\chi I_{coll}[f_\sigma]}
\end{equation}
where the average is taken both in position and momentum space
\begin{equation}
\brak{\chi}= \int \chi(\er,\pp) f(\er,\pp)\, d\Gamma \biggm/ \int
f(\er,\pp)\, d\Gamma 
\end{equation}
having defined $d\Gamma=d^3r\,d^3p/h^3$. The collisional contribution
can be rewritten in the form
\begin{equation} \label{CollIntav}
\sum_\sigma \brak{\chi  I_{coll}[f_\sigma]}= \frac 1 4 \sum_\sigma
\int \Delta \chi(\er,\pp) I_{coll}\left[f_\sigma(\er,\pp)\right]
d\Gamma \biggm/ \int f(\er,\pp)\, d\Gamma 
\end{equation}
where $\Delta \chi=\chi_1+\chi_2-\chi_{1'}-\chi_{2'}$ with
$\chi_i\equiv \chi(\er,\pp_i)$. If $\chi$ is a quantity conserved
during the elastic collision then integral~(\ref{CollIntav}) is zero.
This happens if $\chi=a(\er)+{\mathbf{b}}(\er)\cdot \pp + c(\er)p^2$
with $a$, $\mathbf{b}$ and $c$ arbitrary functions of the position.

As shown in \cite{DavidFrancesca} the dynamical evolution of the
$m_z=0$ quadrupole oscillation in an axially symmetric trapping
potential is described by a system of six coupled equations that we
rewrite here:
\begin{equation} \label{set}
\begin{split}
  &\partial_t \langle \chi_1 \rangle -2 \langle \chi_3 \rangle=0\\
  &\partial_t \langle \chi_2 \rangle -2 \langle \chi_4 \rangle=0 \\
  &\partial_t \langle \chi_3 \rangle -\langle \chi_5 \rangle +\frac{2
    \omega^2_\perp +\omega^2_z} 3 \langle \chi_1 \rangle + \frac
  {\omega^2_z-\omega^2_\perp} 3 \langle
  \chi_2 \rangle=0\\
  &\partial_t \langle \chi_4 \rangle-\langle \chi_6 \rangle + \frac {2
    \omega^2_z-2\omega^2_\perp} 3 \langle \chi_1 \rangle + \frac
  {\omega^2_\perp +2\omega^2_z} 3
  \langle \chi_2\rangle=0 \\
  &\partial_t \langle \chi_5 \rangle+ \frac
  {2\omega^2_z+4\omega^2_\perp} 3 \langle \chi_3 \rangle +\frac
  {2\omega^2_z-2\omega^2_\perp} 3 \langle \chi_4 \rangle =0\\
  &\partial_t \langle \chi_6 \rangle +\frac
  {4\omega^2_z-4\omega^2_\perp} 3 \langle \chi_3 \rangle +\frac {4
    \omega^2_z +2\omega^2_\perp} 3 \langle \chi_4 \rangle =\sum_\sigma
  \langle \chi_6 I_{coll}[f_\sigma] \rangle.
\end{split}
\end{equation}
The $\chi_i$ quantities are defined as
\begin{align}
\chi_1 &=r^2 &
\chi_2 &=2z^2-r_\perp^2 \notag \\
\chi_3 &=\er\cdot\vu &
\chi_4 &=2zv_z-\er_\perp\cdot\vu_\perp \\
\chi_5 &= v^2 &
\chi_6 &=2v_z^2-v_\perp^2 \notag
\end{align}
where $\vu=\pp/m$ and $\er_\perp$, $\vu_\perp$ are respectively the
projections of the $\er$ and $\vu$ vectors in the $xy$ plane. The $m_z=0$
quadrupole oscillation is 
coupled to the monopole one unless the trap is spherical
($\omega_z= \omega_\perp$). The
$\chi_6$ combination is the only one for which the collisional
contribution does not vanishes. To describe the $m_z=2$ quadrupole
oscillation one should instead introduce the mean values of $xy$,
$v_xy+v_yx$ and $v_x v_y$. We shall not consider here this mode.

Solving system~(\ref{set}) is still a hard task because a full
solution of the Boltzmann equation is needed to actually calculate the
average of the collisional contribution. What can be done is to
linearize the distribution function assuming a specific form for its
deviation from equilibrium, finding a variational estimate for
$f(\er,\pp)$. By recalling that in the collisionless regime one can
have solutions with anisotropic equilibrium velocities, corresponding
to different temperatures in the radial and axial directions, we make
the ansatz
\begin{equation} \label{ansatz}
f_\sigma(\er,\pp)=f_0(\er,(1+\alpha_\perp)^{1/2}\pp_\perp,
(1+\alpha_z)^{1/2}p_z) 
\end{equation}
where $f_0$ is the equilibrium distribution function. The proposed
anisotropy in momentum space is of quadrupole type if one chooses
$\alpha_\perp=-\alpha_z/2$. In this case the normalization is
preserved to first order in the deformation parameters. Deformations in
the real space do not need to be taken into account in
ansatz~(\ref{ansatz}) because, after linearization, they do not
contribute to the integral as a consequence of the invariance property
of the collisional integral already discussed.

The linearization of $f_\sigma$ around $\alpha_\perp=\alpha_z=0$ leads to
\begin{equation}
f_\sigma \simeq f_0-\left(\frac {\alpha_\perp \beta p_\perp^2}{2m}+
\frac{ \alpha_z \beta p_z^2}{2m}\right) f_0(1-f_0)
\end{equation}
with $\beta=(k_BT)^{-1}$ and, using the equilibrium identity
$(1-f_0(\pp_1))(1-f_0(\pp_2))f_0(\pp_1')f_0(\pp_2')=f_0(\pp_1)
f_0(\pp_2)(1-f_0(\pp_1'))(1-f_0(\pp_2'))$, we find the following
expression, to first order in the alpha's, for the collisional term:
\begin{multline} \label{firstorder}
\sum_\sigma \brak {\chi_6 I_{coll}[f_\sigma]} \simeq  
\frac{\sigma_f\beta}{16\pi h^6 m N} \int d^3r d^3p_1 d^3p_2 d\Omega
\Delta\chi_6 |\vu_1-\vu_2| \times \\ \left( {\alpha_\perp  \Delta p_\perp^2} +
  {\alpha_z \Delta p_z^2} \right) f_0(\pp_1)f_0(\pp_2)(1-f_0(\pp_1'))
(1-f_0(\pp_2')). 
\end{multline}
Carrying out the calculation (see Appendix) shows that the
collisional contribution is proportional to the mean value of
$\brak{\chi_6}$ so that one can write
\begin{equation} \label{proportionality}
\sum_\sigma \brak{\chi_6 I_{coll}[f_\sigma]}= -\frac {\brak{\chi_6}}{\tau_Q}
\end{equation}
defining the temperature dependent relaxation time for the quadrupole
oscillation.
The dimensionless quantity $\omega_{ho}\tau_Q$ can be written in the
form
\begin{equation} \label{omegatauf}
\frac 1 {\omega_{ho}\tau_Q}=\frac {4}{5\sqrt[3] 3 \pi} \left( N^{1/3}
    \frac a {a_{ho}}\right)^2 F_Q\left( \frac T {T_F} \right) 
\end{equation}
where $a_{ho}=[\hbar/(m\omega_{ho})]^{1/2}$ is the harmonic oscillator
length and $F_Q(t)$ (see Appendix) is an universal function
of the reduced temperature $t=T/T_F$, showed in Figure~\ref{fig:FQ}, which
contains the effect of Fermi statistics. Numerical coefficients have
been chosen so that $F_Q$ has the
limiting form $1/t$ for $t\gg 1$ (classical regime).
In the fully degenerate regime the integral defining $F_Q$ is
analytical and the result 
\begin{equation}\label{FQquantum}
F_Q(t) = 8 \pi^2 t^2
\end{equation}
is found. The $T^{-2}$ behaviour for the relaxation time is typical of
degenerate Fermi systems \cite{Pines,Baym}. We recall that
result~(\ref{FQquantum}) holds above the BCS phase transition.
Conversely we point out explicitly that result~(\ref{omegatauf}) in the
classical limit is correct only as long as the approximations that led
to the collisional integral~(\ref{CollInt}) remain valid. The truly
classical system is the one for which all partial waves are taken into
account in the two body collision, thereby restoring the collisions
between up-up and down-down particles, no longer suppressed by
(quantum) antisymmetry requirements. The classical limit of
equation~(\ref{omegatauf}) instead keeps on having the quantal
characteristic of up-down $s$-wave scattering only. However $p$-wave
collisions in real experiments become important only at temperatures
much higher than the Fermi temperature \cite{DeMarco} and it is safe
to describe in such a way our system. We finally observe that the
relaxation time depends on the square of the scattering length $a$
and, consequently, it is not sensitive to its sign. Mean field
corrections, that have been neglected in the Boltzmann equation,
do depend on the sign of the interaction \cite{SD}.

Using result~(\ref{proportionality}) the system of
equations~(\ref{set}) is closed and the dispersion law for the
collective oscillations can be obtained. Before considering the
oscillation frequencies we address our attention to the number of
collisions that take place in the gas by calculating the collision rate
\begin{equation} \label{rate}
\gamma= \frac {2\sigma_f} {4 \pi Nh^6} \int d^3rd^3p_1d^3p_2 d\Omega\,
 | \vu_1 -\vu_2| f_{1 \uparrow} f_{2 \downarrow}
 (1-f_{1'\uparrow})(1-f_{2'  \downarrow}).
\end{equation}
The factor $2$ in equation~(\ref{rate}) takes into account the presence of
the two components. By treating integral~(\ref{rate}) in the very same
way of the collisional contribution~(\ref{omegatauf}) we find the
following expression for the rate at equilibrium (put in adimensional
form): 
\begin{equation} \label{gammaf}
\frac \gamma {\omega_{ho}} \equiv \frac 1 {\omega_{ho}\tau_\gamma} =
\frac 1 {\sqrt[3] 3 \pi} \left(N^{1/3}\frac a {a_{ho}} 
\right)^2 F_\gamma\left( \frac T {T_F} \right)
\end{equation}
where again the coefficients of the function $F_\gamma$ are chosen so
that it has the $1/t$ classical behaviour.
In the degenerate regime one instead finds the analytical result
\begin{equation} \label{quantgamma}
\frac 1 {\omega_{ho}\tau_\gamma} \underset{T \to 0}{\to} 3^{2/3}\, 8\pi
\left(N^{1/3}\frac a {a_{ho}} \right)^2 \left( \frac T {T_F} \right)^3.
\end{equation}
Result~(\ref{quantgamma}) explicitly shows the importance of using the
relaxation time $\tau_Q$ instead of the inverse of the
collision rate $\tau_\gamma$: while in the classical regime they are
proportional, 
their ratio being $(\tau_Q/\tau_\gamma)_{class.}=5/4$,
in the degenerate regime their temperature dependence is different.
The cubic dependence on temperature of the collision rate in the
quantum degenerate regime is not a consequence of the trapping
but holds also in the uniform system where the same type of analytical
calculations can be carried on as well. 
\section{Comparison with the Bose case} \label{sec3}
The absence of proportionality between collision rate and relaxation
time in a non condensed Bose gas was pointed out by Lopez-Arias and
Smerzi \cite{Smerzi} who numerically calculated the relaxation time
$\tau_Q$ for the uniform gas. The method of averages allows to
calculate $\tau_Q$ 
for the trapped Bose gas by considering the same
ansatz~(\ref{ansatz}) used in the Fermi case. We now have only one
spin component and thus 
the spin index is absent. The cross section is $\sigma_b=8\pi a^2$
because the scattering is between indistinguishable particles and the
collisional integral reads
\begin{multline} \label{CollIntBose}
I_{coll}[f]= \frac {\sigma_b} {4 \pi h^3
  m} \int
d^3p_2\,d\Omega\, 
|\pp_1-\pp_2| \times \\ 
\left[ (1 + f(\pp_1))(1
  +f(\pp_2))f(\pp'_1)f(\pp'_2)-f(\pp_1)f(\pp_2)(1 +
  f(\pp'_1))(1+f(\pp'_2)) \right]. 
\end{multline}
The $(1+f)$ factors in the above equation account for the increase
of probability of a scattering event in an already occupied final state,
typical of Bose systems. 

The expression for the quadrupolar relaxation time $\tau_Q^b$, analog
to~(\ref{omegatauf}) and in agreement with \cite{Kavoulakis}, is given by
\begin{equation} \label{omegataub}
\frac 1 {\omega_{ho}\tau_Q^b}= \frac {16 \sqrt[3]{\zeta(3)}}{5\pi}
\left( N^{1/3} \frac a {a_{ho}} \right)^2 G_Q\left(\frac T {T_c}
  \right)
\end{equation}
and the one for the collision rate, analog to~(\ref{gammaf}), reads
instead
\begin{equation} \label{Ggamma0}
\frac {\gamma_b}{\omega_{ho}}\equiv \frac 1 {\omega_{ho}\tau_{\gamma_b}}=
\frac{4 \sqrt[3]{\zeta(3)}} \pi \left
  ( N^{1/3} \frac a {a_{ho}} \right)^2 G_{\gamma_b}\left(\frac T {T_c}
  \right).
\end{equation}
The functions $G_Q$ and $G_{\gamma_b}$ are plotted in
Figure~\ref{fig:G} and their expressions are given in
Appendix. Both functions behave as $T_c/T$ in the
classical limit. Comparison of the two equations shows again the
limiting ratio $(\tau_Q^b/\tau_{\gamma_b})_{class.}= 5/4$
\cite{DavidFrancesca}. By using an improved trial function the authors
of \cite{Kavoulakis} have shown that in the classical limit the ratio
changes by only $1 \%$, proving the validity of ansatz~(\ref{ansatz}).

The quantum behaviour of $\tau^b_Q$ and $\tau_{\gamma_b}$ is opposite
to the behaviour of the corresponding fermionic quantities. Quantum
statistical effects enhance the number of collisions taking place in
the trapped gas and the rate $\gamma_b$ is seen to increase with
respect to the classical trend. The integral $G_{\gamma_b}$ of
equation~(\ref{Ggamma0}) (see Appendix) is well behaved as $\beta\mu
\to 0^-$ and reaches a finite value, differently from the uniform case
\cite{Smerzi}, where the collision rate diverges at the critical
temperature, driving to zero the time between collisions. This
divergence has also been found by Nikuni \emph{et al.} \cite{Nikuni},
in the framework of a quantum Boltzmann equation coupled with an
equation for the condensate, as $T_c$ is approached both from above
and from below. 
In the non-uniform case instead $\tau_{\gamma_b}$ does not
vanishes at the transition. To understand this it is useful to recall that
$\tau_{\gamma_b}$, as well as $\tau^b_Q$ and their fermionic
counterparts, can be regarded as the average over the whole sample of
a local collision (relaxation) time. It is then natural to find a
weakening of the divergences exhibited by uniform systems.
Figure~\ref{fig:G} shows that the quadrupolar relaxation rate
$(\tau^b_Q)^{-1}$, similarly to $\gamma_b$, increases its value with
respect to the classical value but in a less stronger way as compared
to the increase of the collision rate. This witnesses the loss of
efficiency of the collisions near the transition temperature: the
number of collisions gets larger, due to the Bose factors $(1+f)$, but
the relaxation rate does not increases its value proportionally
\cite{errata}. This
feature was pointed out also in \cite{Smerzi} where the authors even
find a diverging behaviour for the relaxation time of the quadrupolar
oscillation. To obtain a diverging behaviour in our trapped case one
should observe the vanishing of the function $G_Q$ in
equation~(\ref{omegataub}) as $T \to T_c^+$. This is not the case, as
the integral is positive defined and increases its value near the
critical temperature. 
As final remark we observe that the (linearized)
Boltzmann equation is not accurate in describing the system at the
Bose--Einstein phase transition because it completely neglects
fluctuations and coherence phenomena so that one should not look at the
quantum limit $\beta \mu = 0$ as a quantitative prediction. Conversely
the behaviour of $G_Q$ and $G_{\gamma_b}$ as $T \to T_c^+$ is physically
relevant. The coupled dynamics of the condensate and the thermal
component in the trapped case below $T_c$ can be found in
\cite{Zaremba} where the normal component and the Bose condensed
fraction are taken into account in the framework of a quantum
Boltzmann equation coupled with a generalized Gross--Pitaevskii
equation. Finally two examples of theories that also deal with the
issue of the condensate formation, within formalism derived from
quantum optics and from field theory, can be found in
\cite{Gardiner} and \cite{Stoof2}. 
\section{Quadrupole oscillations in the Fermi gas} \label{sec4}
We now consider the frequency and the damping of the quadrupolar
collective oscillation that can be derived from the set of
equations~(\ref{set}) together with~(\ref{proportionality}). By
looking for solution of oscillatory type one finds the dispersion
relation by imposing the determinant of the system to be zero. The
dispersion relation coincides with the one already derived for the
classical gas \cite{DavidFrancesca}, as quantum statistical effects
are kept into account in the relaxation time $\tau_Q$. One gets 
\begin{equation}\label{displaw}
\left( \omega^2-4\omega_z^2\right) \left(\omega^2 -4\omega^2_\perp
\right) - \frac i {\omega\tau_Q} \left( \omega^4 - \frac 2 3 \omega^2
  \left( 5 \omega^2_\perp + 4\omega^2_z\right) + 8 \omega^2_\perp
  \omega^2_z \right) =0.
\end{equation}
The first term of~(\ref{displaw}) corresponds to the dispersion law
for the pure collisionless regime ($\omega\tau_Q \rightarrow \infty$).
In this case the eigenfrequencies coincide with the ones predicted by
the single particle harmonic oscillator Hamiltonian: $\omega_{CL}=2\,
\omega_{\perp,z}$. The total Hamiltonian is in fact the
sum of single particle Hamiltonians and interactions are completely
neglected. In this case the concept of collective oscillation actually
looses his meaning.
On the other hand the term multiplying
$i/(\omega\tau_Q)$ refers to the pure hydrodynamical regime
($\omega\tau_Q \rightarrow 0$) and the dispersion relation gives the
result \cite{GriffinWuS} $\omega^2_{HD}/\omega_\perp^2= (4\lambda^2+5\pm
\sqrt{16\lambda^4-32\lambda^2+25}) /3$ where
$\lambda=\omega_z/\omega_\perp$ is the anisotropy parameter. Putting
$\lambda=1$ gives two frequencies $\omega_{HD}^M=2\,\omega_{ho}$ and
$\omega_{HD}^Q= \sqrt 2\, \omega_{ho}$ corresponding to the monopole and
quadrupole mode respectively (as observed, system~(\ref{set})
decouples in two sub-sets). The relevant limits for deformed traps
$\lambda \ll 1$ and $\lambda \gg 1$ can be obtained as well
\cite{DavidFrancesca,AlKawaja}.  When the relaxation time $\tau_Q$ is
of the order of the oscillation frequency one is in an intermediate
regime and equation~(\ref{displaw}) is solved numerically for
$\omega=\omega_r + i \Gamma$ yielding the frequency $\omega_r$ and the
damping $\Gamma$ of the oscillation. As observed in \cite{Kavoulakis2}
in the Bose gas case, the notion of hydrodynamic regime in trapped
gases fails at the boundary of the trap, where the gas is rarefied and
collisions are rare. The average performed over all the sample in
calculating the relaxation time $\tau_Q$ takes care of this effect.
Indeed the possibility of an unified description of the collisionless
and of the hydrodynamic regimes is the useful feature of the presented
model.

\begin{figure}[htbp]
  \begin{center}
    \input{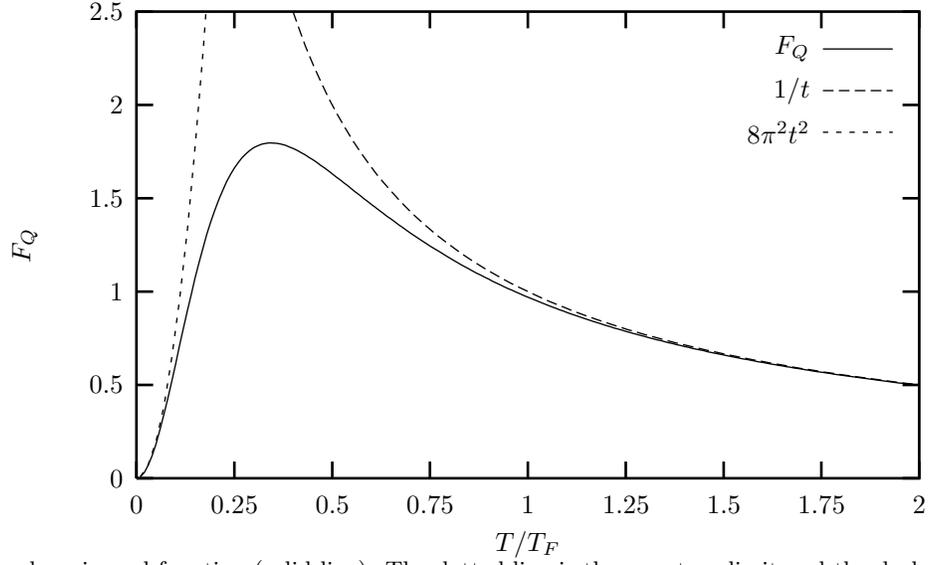}
    \caption{Quadrupole universal function (solid line). The dotted
      line is the quantum limit and the dashed line is the classical
      prediction.}
    \label{fig:FQ}
  \end{center}
\end{figure}

By looking at the temperature dependence of the relaxation rate,
depicted in Figure~\ref{fig:FQ}, it is clear that the gas will be in
the collisionless regime both at low and high temperatures. At low
temperatures the Pauli principle reduces the number of available final
states for the scattering process and the relaxation rate tends to
zero as $(T/T_F)^2$. At high temperatures, as noted previously,
the relaxation rate is proportional to the collision rate and
decreases as $T_F/T$. This is basically a density effect: for fixed
number of particles a Boltzmann gas gets more and more rarefied as
temperature increases. In the intermediate region there is room for an
increase of $\tau_Q^{-1}$. Whether the relaxation time $\tau_Q$ will reach
a sufficiently small value to bring the system in the hydrodynamic
regime depends on the combination $(N^{1/3}a/a_{ho})^2$ of the 
number of particles, the scattering length and the trap parameters of
equation~(\ref{omegatauf}) as the function $F_Q$ depends on the
reduced temperature only. The maximum of $F_Q$ is $\sim 1.8$ at
$T/T_F\simeq 0.35$ so that a
minimum value of $\omega_{ho}\tau_Q=3.15/(N^{1/3}a/a_{ho})^2$ is
reached. Large numbers of trapped atoms as well as large scattering lengths
should allow the observation of the hydrodynamic regime. In the
following we will consider explicitly the relevant experimental cases
of $^{40}$K \cite{Jin} and $^6$Li \cite{Paris}.

The crossover from collisionless to the
hydrodynamic regime is most striking in the case of the spin dipole
oscillation \cite{SD} where the dispersion law is simpler
than~(\ref{displaw}) and allows to discriminate analytically between a
damped oscillatory 
behaviour and an overdamped one. In the
case of spin dipole oscillations one can find overdamping because the
spin current $\vu_\uparrow-\vu_\downarrow$ is not
conserved during collisions and this leads to a purely diffusive mode
in the collision dominated regime. In the case of quadrupole oscillations
instead both the collisionless and the hydrodynamic regimes support an
oscillation and the actual regime can be deduced from the value of the
real part of the oscillation frequency which passes, \emph{e.g.} in the
spherical case, from $2\,\omega_{ho}$ to 
$\sqrt 2\, \omega_{ho}$.
The spectrum of a trapped Fermi gas in the hydrodynamic regime has been
studied independently by Amoruso \emph{et al.}~\cite{Amoruso} and by
Bruun and Clark \cite{Bruun2} for $T \ll T_F$ and $T \gg T_F$. 
The frequency of the quadrupolar
oscillation clearly coincides with result~(\ref{displaw}) in the
$\omega \tau_Q \to 0$ limit. However we observe that in the
limits of low and high temperature the value of the oscillation
frequency will not be the hydrodynamic but the collisionless one. 
This even supplies a possible signature for the
superfluid transition since the value of the quadrupole frequency for a
superfluid Fermi gas coincides with the hydrodynamic value
\cite{superfluid-oscill} and at the calculated temperature for the BCS phase
transition \cite{Kagan,Stoof} $\tau_Q$ is large and the system is in
the collisionless regime. 
So for large number of particles and for large values of
$|a|/a_{ho}$ one should observe, as a function of the decreasing
temperature, the sequence of collisionless -- hydrodynamic --
collisionless -- hydrodynamic value for the real part of the
quadrupole frequency and 
a corresponding structure of local maxima and minima in the damping
rate of the oscillation.

To summarize this discussion we show in Figure~\ref{fig:K} the real
and the imaginary part of the quadrupolar oscillation of a spherical
($\lambda=1$) cloud of $^{40}$K atoms. The scattering length is
\cite{Jin} $a=157\,a_0$ with $a_0$ the Bohr radius. We take the ratio
$a/a_{ho} \sim 10^{-3}$ and $N=10^6$. The minimum value of
$\omega_{ho}\tau_Q$ is then $\sim 350$ and the system is always in the
collisionless regime. In the upper panel the real part of the
frequency (solid line) is compared with the frequency of a Boltzmann
gas (dashed line) which, at
low temperatures, is always in the hydrodynamic regime. This
emphasizes the role of Fermi statistics in the low temperature regime
of the trapped gas. The lower panel shows the damping of the
oscillation. The two gases exhibit zero damping in the pure
collisionless regime at 
high temperature. At zero temperature both the Boltzmann gas and the
Fermi gas have zero damping but the former is in the pure hydrodynamic 
regime while the latter is in the pure collisionless one because of
Pauli blocking. Inspection
of the damping rate gives then another clear signature of quantum degeneracy
as it departs significantly from the classical value.

Figure~\ref{fig:Li} shows instead the real (upper panel) and the
imaginary (lower panel) part of the quadrupolar oscillation frequency
for $N=10^6$ atoms of $^6$Li, that have a resonant value of the
scattering length $a=-2160\,a_0$, in a spherical trap.
Due to the lighter mass with respect to potassium the ratio
$|a|/a_{ho}$ is $\sim 5\times 10^{-2}$ and a minimum value of
$\omega_{ho}\tau_Q \sim 0.14$ is found. Lithium can thus be found in the
hydrodynamic regime in the region around the maximum of the function $F_Q$
(Figure~\ref{fig:FQ}). Indeed the real part of the frequency bends towards the
hydrodynamic value $\sqrt 2\,\omega_{ho}$ as long as the Fermi
statistics allows a sufficient collisional activity. Correspondingly
the damping has a relative minimum. One should point out that if the
system had reached the pure hydrodynamical regime at $T\simeq 0.35\,
T_F$ the damping would have then been exactly zero. As the temperature
is further lowered Pauli principle forces the system to become again
collisionless, the real part raises again towards the value $2\,
\omega_{ho}$ while the damping exhibit a local maximum and then falls
to zero once the complete collisionless regime is reached. 

The two proposed examples involve the measured scattering lengths for
potassium and lithium. Different predictions for the frequency and the
damping can be obtained for different values of the number of trapped
atoms, as well 
as of the trapping parameters and scattering length 
through the use of the definition~(\ref{omegatauf}) of the relaxation
time together with the universal function plotted in
Figure~\ref{fig:FQ}. A way to reach the hydrodynamic regime also in
the case of the potassium isotope could be the change of the value of
its scattering length with external fields, as already experimentally
observed in alkali atoms \cite{Inouye}. 
\section{Conclusions} \label{sec5}
We have studied the collective frequencies and the damping of a
trapped normal Fermi system through the use of a linearized Boltzmann
equation, a method that gives a variational estimate for the
relaxation time $\tau_Q$ relative to the coupled monopole-quadrupole
oscillations. The effects of Fermi statistics are included in the
relaxation time $\tau_Q$ that diverges at zero temperature as a
consequence of the Pauli principle and at classical temperatures $T
> T_F$ as a consequence of the diluitness of the gas. In both cases
the collisionless regime for the collective oscillation is reached. At
intermediate temperatures we have found the possibility of a
hydrodynamical behaviour of the quadrupole mode for the $^6$Li
isotope, due to his large scattering length, for reasonable number of
trapped particles. We have also compared the relaxation rate $\tau_Q^{-1}$
with the collision rate $\gamma$, finding analytically
different behaviours at low temperatures. This emphasizes the need of
distinguishing $\tau_Q$ from $\gamma^{-1}$. The role of quantum
statistics has also been considered in the case of a Bose gas and the
difference between the collision rate and the relaxation rate have
been pointed out also in this case. Future developments will
concentrate on non symmetric configurations, that may be relevant from
the experimental point of view, as well as on the study of the damping
of collective oscillations in mixed systems.

Useful discussions with A. Smerzi and H. T. C. Stoof are
acknowledged. I wish to thank S. Stringari for critically reading the
manuscript.
%
%
\appendix \section{Calculation of the collisional integral} \label{app}
We first introduce the center of mass variables
\begin{equation}  \label{change}
\begin{split}
\vu_1= \Vu + \frac \vu 2 \qquad & \qquad \vu'_1= \Vu + \frac {\vu'} 2 \\
\vu_2= \Vu - \frac \vu 2 \qquad & \qquad \vu'_2= \Vu - \frac {\vu'} 2 
\end{split}
\end{equation}
where $|\vu|=|\vu'|=v$ for momentum conservation. The Jacobian
associated to transformation~(\ref{change}) is
one. In these variables one has
\begin{equation}
\begin{split}
\Delta v_\perp^2&=-\frac 1 2 \left(v_z^2-v_z^{\prime\, 2} \right)\\
\Delta v_z^2&=\frac 1 2 \left(v_z^2-v_z^{\prime\, 2} \right)
\end{split}
\end{equation}
and  $\Delta \chi_6=2\Delta v_z^2-\Delta v_\perp^2$.
The product $f_0(\vu_1)f_0(\vu_2)(1-f_0(\vu_1'))(1-f_0(\vu_2'))$ can
be written in the useful form
\begin{equation}
f_0(\vu_1)f_0(\vu_2)(1-f_0(\vu_1'))(1-f_0(\vu_2'))= 
\frac 1 4 \frac 1 {\cosh(\nu)+\cosh(\gamma \Vu\cdot \vu)} 
\frac 1 {\cosh(\nu)+\cosh(\gamma \Vu\cdot \vu')} 
\end{equation}
where $\nu=\beta(m(V^2+v^2/4)/2+m\omega_\perp^2(x^2+y^2+\lambda^2z^2)/2-\mu)$
and $\gamma=\beta m/2$. We then have to linear order in
$\alpha_\perp$, $\alpha_z$
\begin{equation} \label{chi6icoll}
\sum_\sigma \brak{\chi_6 I_{coll}[f_\sigma]}=
\frac {3 \sigma_f \beta m^7 \delta \alpha}{2^8 \pi h^6 N} \times I
\end{equation}
with $\delta \alpha=\alpha_z-\alpha_\perp$ and
\begin{equation}\label{integral}
I=\int
d^3rd^3V d^3v d\Omega\,v \left(v_z^2-v_z^{\prime\, 2} \right)^2
\frac 1 {\cosh(\nu)+\cosh(\gamma \Vu\cdot \vu)} \cdot
\frac 1 {\cosh(\nu)+\cosh(\gamma \Vu\cdot \vu')} .
\end{equation}

Integral~(\ref{integral}) can be carried on as follows:
introduce a reference frame $(\hat {\mathbf{a}},\hat {\mathbf{b}},\hat
{\mathbf{c}})$ oriented such that $\Vu=V \hat {\mathbf{a}}$ and $\hat
{\mathbf{z}}$ is in the plane generated by $\hat {\mathbf{a}}$ and $\hat
{\mathbf{b}}$. We then have the following expressions:
\begin{align} \label{vz}
v_z&=v_a (\hat {\mathbf{a}} \cdot \hat {\mathbf{z}})+ v_b (\hat {\mathbf{b}}
\cdot \hat {\mathbf{z}}) = v \cos(\theta)(\hat {\mathbf{a}} \cdot \hat
{\mathbf{z}}) +v \sin(\theta)\cos(\phi)(\hat {\mathbf{b}}
\cdot \hat {\mathbf{z}}) \\ \label{vzprimo}
v'_z&= v'_a (\hat {\mathbf{a}} \cdot \hat {\mathbf{z}})+ v'_b (\hat
{\mathbf{b}} 
\cdot \hat {\mathbf{z}}) = v \cos(\theta')(\hat {\mathbf{a}} \cdot \hat
{\mathbf{z}}) +v \sin(\theta')\cos(\phi')(\hat {\mathbf{b}}
\cdot \hat {\mathbf{z}})
\end{align}
and $\Vu \cdot \vu^{(\prime)}=
Vv\cos(\theta^{(\prime)})$. Integrations in $d\phi$ and
$d\phi'$ are immediate. One can then integrate in $d\theta$ and
$d\theta'$ using the fact that $\cosh(x)$ is an even function to end
with the following intermediate form of the integral 
\begin{equation}\label{intquad2}
I= \frac {2^7 \pi^4} {5\lambda} \int r^2dr V^2dV v^7dv\int_{-1}^1 dx
dy \left(\frac 1 3 + \frac 2 3 x^2-  x^2y^2 \right)
\frac 1 {\cosh(\nu)+\cosh(\gamma Vvx)}\cdot \frac 1
{\cosh(\nu)+\cosh(\gamma Vvy)}.
\end{equation}
A final change of variables to polar dimensionless coordinates leads
to the final form of $I$ which has to be integrated numerically except
in the low and high temperature limits:
\begin{multline}\label{intquad3}
I=\frac {2^{21} \pi^4}{5\omega^3_{ho}} \frac 1 {(\beta m)^7} \int_0^\infty
\rho^6d\rho \int_0^{\frac \pi 2} d\vartheta d\varphi \cos^2(\vartheta)
\sin^{10}(\vartheta) \sin^7(\varphi) \cos^2(\varphi) \times \\
\int_{-1}^1 d\eta d\xi \left( \frac 1 3 + \frac 2 3 \eta^2 -\eta^2
  \xi^2 \right)
\frac 1{\cosh(\rho-\beta\mu)+\cosh(\rho\sin^2(\vartheta)
  \sin(2\varphi)\eta)} \times\\ \frac 1
{\cosh(\rho-\beta\mu)+\cosh(\rho\sin^2(\vartheta)
  \sin(2\varphi)\xi)}.
\end{multline}

The value of $\brak{\chi_6}$ can be calculated to be, to first order
in $\delta \alpha$
\begin{equation} \label{chi6}
\brak{\chi_6}=-\frac {4 \delta \alpha} {N m \beta^4 (\hbar
  \omega_{ho})^3} f_4(z)
\end{equation}
and the common proportionality of~(\ref{integral}) and (\ref{chi6}) on
$\delta\alpha$ allows to write equation~(\ref{proportionality}) of the
text. Using~(\ref{chi6}) and (\ref{intquad3}) one builds the $F_Q$
function as
\begin{multline}
F_Q\left(\frac T {T_F} \right)= \frac {9\, 2^5}{f_4(z)}
\left(\frac T {T_F} \right)^2 \int_0^\infty
\rho^6d\rho \int_0^{\frac \pi 2} d\vartheta d\varphi \cos^2(\vartheta)
\sin^{10}(\vartheta) \sin^7(\varphi) \cos^2(\varphi) \times \\
\int_{-1}^1 d\eta d\xi \left( \frac 1 3 + \frac 2 3 \eta^2 -\eta^2
  \xi^2 \right)
\frac 1{\cosh(\rho-\beta\mu)+\cosh(\rho\sin^2(\vartheta)
  \sin(2\varphi)\eta)} \times\\ \frac 1
{\cosh(\rho-\beta\mu)+\cosh(\rho\sin^2(\vartheta)
  \sin(2\varphi)\xi)}.
\end{multline}
In the classical limit $\beta\mu \to - \infty$ and
the denominators such as $\cosh((\rho-\beta\mu)+\cosh(\rho\sin^2(\vartheta)
  \sin(2\varphi)\xi)$ can be approximated with
$\exp(\rho-\beta\mu)/2$. In the quantum degenerate regime instead
$\beta \mu \to \infty$ and, after changing variables to
$\rho'=(\rho-\beta \mu)$, $\xi'=(\rho'+\beta\mu)
\sin^2(\vartheta)\sin(2\varphi)\xi$ and $\eta'=(\rho'+\beta\mu)
\sin^2(\vartheta)\sin(2\varphi)\eta$ one has integrals over the whole
real axis that can be done analytically.

The same type of calculations can be done for the collision rate as
well as for the corresponding bosonic expressions of $\tau_Q^b$ and
$\tau_{\gamma_b}$. The formula for $F_\gamma$ is:
\begin{multline}
F_\gamma \left(\frac T {T_F} \right)=\frac {9\, 2^7} \pi 
\left(\frac T {T_F} \right)^5 \int_0^\infty d\rho
\,\frac{\rho^2}{\sinh^2(\rho-\beta\mu)} 
\int_0^{\pi/2} d \theta \,\sin^2(2\theta)\int_0^{\pi/2} d\phi\, \cos(\phi)
\,\times
\\ \mathrm{atanh}^2 \left( \tanh
      \left( (\rho-\beta\mu)/2 \right) \tanh \left( \rho
      \sin^2 (\theta) \sin(2\phi)/2 \right) \right).
\end{multline}

In the Bose case one finds the Bose
function $g_4(z)$ in place of the Fermi function $f_4(z)$ and the
expressions for the bosonic quadrupolar 
relaxation function $G_Q$ and for the bosonic rate function
$G_{\gamma_b}$ are
\begin{multline} \label{GQ}
G_Q\left(\frac T {T_c} \right)= \frac {48}{\zeta(3)\pi g_4(z)}
\left(\frac T {T_c} \right)^2 \int_0^\infty
\rho^6d\rho \int_0^{\frac \pi 2} d\vartheta d\varphi \cos^2(\vartheta)
\sin^{10}(\vartheta) \sin^7(\varphi) \cos^2(\varphi) \times \\
\int_{-1}^1 d\eta d\xi \left( \frac 1 3 + \frac 2 3 \eta^2 -\eta^2
  \xi^2 \right)
\frac 1{\cosh(\rho-\beta\mu)-\cosh(\rho\sin^2(\vartheta)
  \sin(2\varphi)\eta)} \times\\ \frac 1
{\cosh(\rho-\beta\mu)-\cosh(\rho\sin^2(\vartheta)
  \sin(2\varphi)\xi)},
\end{multline}
\begin{multline} \label{Ggamma}
G_{\gamma_b}\left(\frac T {T_c} \right)=\frac {2^5}{\pi \zeta(3)^2}
\left(\frac T {T_c} \right)^5 \int_0^\infty
\frac{\rho^2\,d\rho}{\sinh^2(\rho-\beta\mu) }
\int_0^{\pi/2}d\theta\,d\phi \, \sin^2(2\theta)\cos(\phi)\times \\
\mathrm{atanh}^2\left(\frac{\tanh(\rho \sin^2(\theta) \sin(2
      \phi)/2)} {\tanh\left((\rho-\beta\mu)/2\right)} \right).
\end{multline}
The classical limit is performed again by taking $\beta\mu \to
-\infty$. Differently from the Fermi case, as the critical
temperature $T_c$ is approached the fugacity tends to one, \emph{i.e.}
$\beta \mu \to 0$. In this limit the integrand in~(\ref{Ggamma})
diverges in the origin as a logarithm but the integral is finite.
This is a consequence of the confinement, since in the uniform
case the collision rate diverges at the critical temperature
\cite{Smerzi,Nikuni}. Integral~(\ref{GQ}) can be worked out analytically to
include the effect of quantum degeneracy to first order in the
fugacity $z=\exp(\beta \mu)$ far from the transition temperature ($t
\gg 1$). We perform the expansion to check
explicitly the equivalence of our expression~(\ref{GQ}) with the one
presented in \cite{Kavoulakis}. By expanding the denominator in the
integral in powers of the fugacity and re-introducing Cartesian
coordinates we obtain 
\begin{multline}
G_Q(t) \simeq \frac{128 t^2 z^2}{\zeta(3)\pi g_4(z)} \int_0^{\infty}
x_3^2 dx_3 x_2^7 dx_2 x_1^2 dx_1 \int_{-1}^1 d\eta d\xi \left( 1+2
  \eta^2 -3\eta^2 \xi^2 \right) \times \\
\left\{\exp[-2(x_1^2+x_2^2+x_3^2)]+2z\exp[-3(x_1^2+x_2^2+x_3^2)] \left
      ( \cosh(2 x_1 x_2 \eta)+\cosh(2 x_1 x_2 \xi) \right) \right\} 
\end{multline}
which gives 
\begin{equation}
G_Q(t)\simeq \frac {t^2 z^2}{\zeta(3)g_4(z)}\left( 1 + \frac 3 8 z \right).
\end{equation}
From the normalization condition $g_3(z)=\zeta(3)/t^3$ one obtains the
correction to the fugacity
\begin{equation}
z\simeq \frac {\zeta(3)} {t^3}-\frac 1 8 \left(\frac {\zeta(3)}
  {t^3}\right)^2
\end{equation}
which, together with the expansion of $g_4(z)\simeq z + z^2/16$ lead
to
\begin{equation}
G_Q(t) \underset{{t \gg 1}}{ =} \frac 1 t \left(1 + \frac 3 {16} \frac
  {\zeta(3)}{t^3}  \right)
\end{equation}
which is the classical limit $1/t$ with the quantum
correction of \cite{Kavoulakis}. A similar expansion can be performed
for the rate function $G_{\gamma_b}$ as well as for the functions
$F_Q$ and $F_\gamma$. We report the result for the $F_Q$ function
which is
\begin{equation}
F_Q(t) \underset{{t \gg 1}}{ =} \frac 1 t \left(1 - \frac 1 {32} \frac
  1{t^3}  \right).
\end{equation}
%
%

\begin{figure}[htbp]
  \begin{center}
    \input{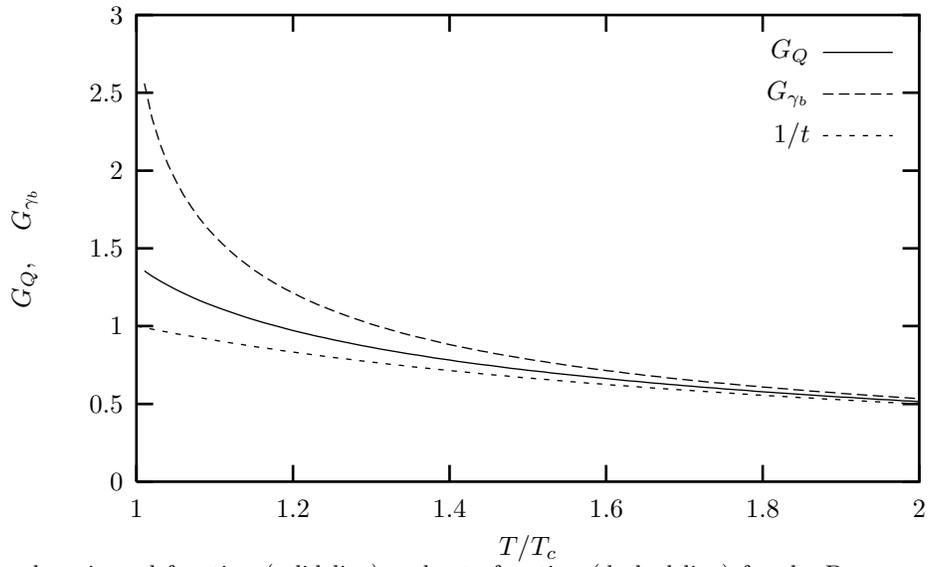}
    \caption{Quadrupole universal function (solid line) and rate
      function (dashed line) for the Bose case as a function of the
      reduced temperature $T/T_c$. The classical limit $1/t$ is shown
      as a dotted line.}
    \label{fig:G}
  \end{center}
\end{figure}
\begin{figure}[htbp]
  \begin{center}
    \input{freqquadK.tex}
    \caption{Real and imaginary part of the quadrupole oscillation
      frequency for $N=10^6$ $^{40}$K atoms (solid lines). The dashed
      lines are the corresponding values for a classical gas.}
    \label{fig:K}
  \end{center}
\end{figure}
\begin{figure}[htbp]
  \begin{center}
    \input{freqquadLi.tex}
    \caption{Real and imaginary part of the quadrupole
      oscillation frequency for $N=10^6$ $^6$Li atoms (solid lines).
      The dashed lines are the corresponding values for a classical
      gas.}  
    \label{fig:Li} 
  \end{center} 
\end{figure}
\end{document}